\input ./style/arxiv-general.cfg
\documentclass[aoas,MSNbibl,nameyear,dvips]{arximspdf}
\makeatletter
   \@ifpackageloaded{graphicx}{}{\usepackage{graphicx}}
\makeatother
\usepackage{url,breakurl}

\doi{10.1214/15-AOAS843}
\volume{9}
\issue{3}
\pubyear{2015}
\firstpage{1328}
\lastpage{1349}
\docsubty{FLA}

\makeatletter
\def\calG{\mathcal{G}}
\def\calL{\mathcal{L}}
\def\calN{\mathcal{N}}
\def\calS{\mathcal{S}}

\newcommand{\E}{\mathrm{E}}

\def\indicator#1{\mathbf{1}_{\{#1\}}}
\def\crps{{\mathrm{CRPS}}}
\def\logit{\mathrm{logit}}
\makeatother

\begin{document}
\begin{frontmatter}

\title{Probabilistic wind speed forecasting on a grid based on ensemble
model output statistics}
\runtitle{Gridded probabilistic wind speed forecasting}

\begin{aug}
\author[A]{\fnms{Michael} \snm{Scheuerer}\corref{}\ead[label=e1]{michael.scheuerer@noaa.gov}\thanksref{t1}}
\and
\author[B]{\fnms{David} \snm{M\"oller}}
\runauthor{M.~Scheuerer and D.~M\"oller}
\thankstext{t1}{Supported in part by the German Federal Ministry of
Education and Research,
in the framework of the extramural research program of Deutscher
Wetterdienst (DWD).}
\affiliation{University of Colorado, Cooperative Institute for
Research in
Environmental Sciences at the NOAA Earth System Research Laboratory and
Ruprecht-Karls-Universit\"{a}t Heidelberg}
\address[A]{NOAA/ESRL\\
Physical Sciences Division\\
University of Colorado\\
325 Broadway, R/PSD1\\
Boulder, Colorado 80305-3337\\
USA\\
\printead{e1}}
\address[B]{Institute of Applied Mathematics\\
Heidelberg University\\
Im Neuenheimer Feld 294\\
D-69120 Heidelberg\\
Germany}
\end{aug}

%
\received{\smonth{2} \syear{2014}}
%
\revised{\smonth{5} \syear{2015}}

%
\begin{abstract}
Probabilistic forecasts of wind speed are important for a wide range of
applications, ranging from operational decision making in connection
with wind power generation to storm warnings, ship routing and
aviation. We
present a statistical method that provides locally calibrated, probabilistic
wind speed forecasts at any desired place within the forecast domain based
on the output of a numerical weather prediction (NWP) model. Three
approaches for wind speed post-processing are proposed, which use either
truncated normal, gamma or truncated logistic distributions to make
probabilistic predictions about future observations conditional on the
forecasts of an ensemble prediction system (EPS). In order to provide
probabilistic forecasts on a grid, predictive distributions that
were calibrated with local wind speed observations need to be interpolated.
We study several interpolation schemes that combine geostatistical methods
with local information on annual mean wind speeds, and evaluate the proposed
methodology with surface wind speed forecasts over Germany from the COSMO-DE
(Consortium for Small-scale Modelling) ensemble prediction system.
\end{abstract}

%
\end{frontmatter}

\section{Introduction}
\label{intro}

The prediction of wind speed over different time scales is one of the
tasks of weather agencies with the widest range of applications. Arguably,
the most important application is wind power forecasting, which is gaining
enormous significance with many countries and regions introducing policies
to increase the use of renewable energy: the European Union is aiming
(by 2020)
to increase the amount of renewable energy to 20\% of the energy
supply, with
wind power playing a key role [\citet
{EUrenewable2008,EUwindenergy2008}]; the
U.S. Department of Energy (DOE) describes a scenario in which wind energy
could provide 20\% of the U.S. electricity demand in 2030
[\citet{DOEreport2008}]; legislation in China declares the usage of renewable
energy a prioritized area in energy development [\citet{ChinaRenewables2005}].
Probabilistic wind power forecasts are most useful, as they permit its optimal
management and trading [\citet{Pinson2013}], and one possibility to
obtain them
is by converting probabilistic forecasts of wind speed to power based on
stochastic power curves [\citet{JeonTaylor2012}].

Accurate forecasts of wind speed are not only required for wind power
prediction, but are crucially important also in connection with severe weather
warnings for the general public. Warnings may be issued either based on wind
speed forecasts directly or based on forecasts of wind gusts, which can be
derived from the former using gust factors [\citet{Durst1960,
ThorarinsdottirJohnson2012}]. Further applications where wind speed forecasts
are required include risk assessment and decision making in aviation, ship
routing, recreational boating and agriculture. Again, it has been argued
that principled risk management should be based on probabilistic forecasts
that take the form of predictive probability distributions for future
quantities or events [\citet{NationalResearchCouncil2006, Gneiting2008}].

To provide probabilistic forecasts with lead times between a few hours
up to
several days, an increasing number of weather centers are running ensemble
prediction systems (EPSs). Instead of a single forecast, several different
forecasts $f_1,f_2,\ldots,f_m$---a so-called ensemble---are
generated, with ensemble members corresponding to model integrations
that differ
in the initial conditions and/or the numerical representation of the atmosphere
[\citet{Palmer2002}]. Combinations of ensemble member forecasts are
often more
accurate than any of these forecasts individually, and their spread provides
useful information on the flow-dependent uncertainty. If the forecasts
$f_1,f_2,\ldots,f_m$ are interpreted as a sample of a predictive distribution,
the corresponding empirical cumulative distribution function (CDF) can be
formed, and probabilistic forecasts can be derived from it. It turns out,
however, that these \textit{raw ensemble forecasts} are often underdispersive
and capture only part of the forecast uncertainty [\citet{HamillColucci1997,
Buizza2005}]. Moreover, forecasts may suffer from systematic biases due to
structural model deficiencies shared among all ensemble members or due to
insufficient resolution. To overcome these deficiencies and provide calibrated,
probabilistic forecasts, methods for statistical post-processing of ensemble
forecasts have been proposed. Here, we focus on approaches that
transform the
ensemble forecasts into a full predictive CDF. These methods are appealing
because one can derive prediction intervals, probabilities of threshold
exceedance, etc. from the predictive CDFs in a consistent way. Furthermore,
for any decision problem that can be expressed in terms of a scoring function
(loss function), an optimal point forecast can be derived from the predictive
distribution using the Bayes rule [\citet{Gneiting2011a}].

The common idea of all methods for statistical post-processing is that
forecast-observation pairs from the past can be used to identify
shortcomings of
the raw ensemble, and generate predictive distributions that do not
suffer from
these shortcomings. Examples of such methods for wind speed ensemble forecasts
include adaptations of the nonhomogeneous Gaussian regression (NGR) approach
[\citet{Gneiting2005}] and adaptations of the Bayesian model averaging (BMA)
technique [\citet{Raftery2005}]. Instead of Gaussian distributions,
\citet{Sloughter2010}, \citet{Courtney2013} and Baran, Nemoda and Hor{\'
{a}}nyi
(\citeyear{Barann2013}) use
gamma distributions as the building block for their predictive BMA densities.
\citet{ThorarinsdottirGneiting2010} use predictive truncated normal
distributions; \citet{LerchThorarinsdottir2013} further extended this approach
and use either truncated normal distributions or generalized extreme value
distributions, depending on whether the forecasts suggest a low or high wind
regime. All of these approaches have been demonstrated to be able to generate
calibrated and sharp predictive distributions, which is the goal in
probabilistic forecasting [\citet{GneitingBalabdaouiRaftery2007}]. They
can be
applied either to stations individually and use only the local wind speed
forecasts and observations as training data, or they can pool data
across the
forecast domain and estimate a single set of model parameters that is valid
on all locations. \citet{ThorarinsdottirGneiting2010} studied both approaches
and found that the local method yields better results than the regional
method, as it allows the post-processing to adapt to local
peculiarities. It
entails, however, a new challenge that none of the above-mentioned articles
have dealt with: when forecasts are desired at locations where no wind speed
measurements are available, either the post-processing parameters or the
parameters of the predictive distributions must be interpolated to those
locations. In operational practice, forecasts are usually provided on a regular
model grid, and the interpolation of local forecasts to this grid is
referred to
as \textit{gridding}. \citet{Kleibera2011} and \citet
{ScheuererBueermann2014} have
proposed procedures for the gridding of BMA- and NGR-based probabilistic
forecasts for temperature. In this paper we will
do the following:
\begin{itemize}
\item compare three different NGR type approaches for probabilistic
wind speed
forecasting based on truncated normal, gamma and truncated logistic
distributions;
\item adapt the model fitting concept by \citet
{ScheuererBueermann2014} of
splitting post-processing parameters into local and regional ones, thus
achieving a good compromise between local adaptivity and parsimony of
the NGR model;
\item study and compare different geostatistical models for the
gridding of
probabilistic wind speed forecasts, placing special emphasis on the adequate
consideration of spatial heterogeneity and small scale variability of observed
wind speeds.
\end{itemize}

After providing some details on the data used in our study in
Section~\ref{sec:2},
some exploratory analysis is performed. We briefly review the NGR type
approach by \citet{ThorarinsdottirGneiting2010} in Section~\ref
{sec:3}, and
propose two alternative methods that use predictive gamma and truncated logistic
distributions, but are otherwise similar. A description of the corresponding
model fitting procedure, in which the continuous ranked probability
score (CRPS)
is minimized, is also given in this section. In Section~\ref{sec:4},
we address
the interpolation problem
mentioned above, propose a geostatistical interpolation scheme that incorporates
information on local annual mean wind speeds, and use this model for obtaining
gridded forecasts. The performance of the different methods with our
data set is
assessed in Section~\ref{sec:5}, and conclusions are drawn about the optimal
training sample size, predictive distribution and interpolation scheme, before
summing up and discussing directions for further extensions.
Mathematical details about the derivation of closed-form expressions
for the
CRPS of gamma and truncated logistic distributions are provided in the
\hyperref[app]{Appendix}.

\section{Data description and exploratory analysis}
\label{sec:2}

We consider surface (10~m) wind forecasts by the COSMO-DE-EPS
(Consortium for
Small-scale Modelling), a multi-analysis and multi-physics ensemble prediction
system based on the high-resolution (2.8 km horizontal grid size) numerical
weather prediction model COSMO-DE [\citet{Baldauf2011}]. The
COSMO-DE-EPS has
been operational at the German Weather Service (DWD) since May 22,
2012. It was
run under the same conditions in a pre-operational phase since 9
December 2010,
consists of $m=20$ ensemble members, covers the area of Germany, and produces
forecasts with lead times up to 21 hours. A new model run is started
every three
hours; we use the one initialized at 0000 UTC and study forecasts at
0600, 1200 and 1800 UTC. The current setup of the lateral
boundary conditions uses forecasts of four different global models,
while five
different (fixed) configurations of the COSMO-DE model are used for the
variation
of model physics [\citet{Gebhardt2011}]. Thus, all 20 ensemble members have
individually distinguishable physical features and are \textit{not
exchangeable}.
The COSMO model uses a rotated spherical coordinate system in order to project
the geographical coordinates to the plane with distortions as small as possible
[\citet{DomsSchaettler2002}, Section~3.3], with $421\times461$
equidistant gridpoints
in longitudinal and latitudinal direction. We adopt this coordinate
system to
calculate horizontal distances in the framework of our post-processing method.

Both raw and post-processed forecasts are verified against surface wind speed
observations (10-minute average wind speed 10~m above the ground) at 286 surface
synoptic observation (SYNOP) stations in Germany. Stations with nonmissing
data on less than 200 days in either 2011 or 2012 have been left out.
The station
at Berlin Alexanderplatz has been left out, too, since the magnitude of the
observations at this site suggests that measurements have actually been
taken at
the top of the Fernsehturm (TV tower), and hence cannot be considered 10~m
wind speeds.
The ensemble forecasts are originally given as the zonal and meridional
component
of 10~m wind vectors, and we take the Euclidean norm of these vectors as the
overall wind speed forecasts and interpolate them to the observation
sites via
bilinear interpolation. In this paper, the aim is to forecast local observations
rather than representative averages over model grid cells, and we neglect
measurement errors and take the wind speed observations as the truth.

The gridded high-resolution (200~m horizontal grid size) data of annual
mean wind
speeds over Germany, which is used as a covariate in our spatial interpolation
scheme in Section~\ref{sec:4}, was also obtained from DWD. It is
constructed based
on measurements at 218 SYNOP stations over Germany during the period
from 1981 to
2000, which were adjusted for obstacles, and gridded using the statistical
wind field model described in the European wind atlas [\citet{EuropeanWindAtlas,
GerthChristoffer1994}]. Values at the station locations and the COSMO-DE model
grid were derived from this high resolution map using bilinear interpolation.

\begin{figure}

\includegraphics{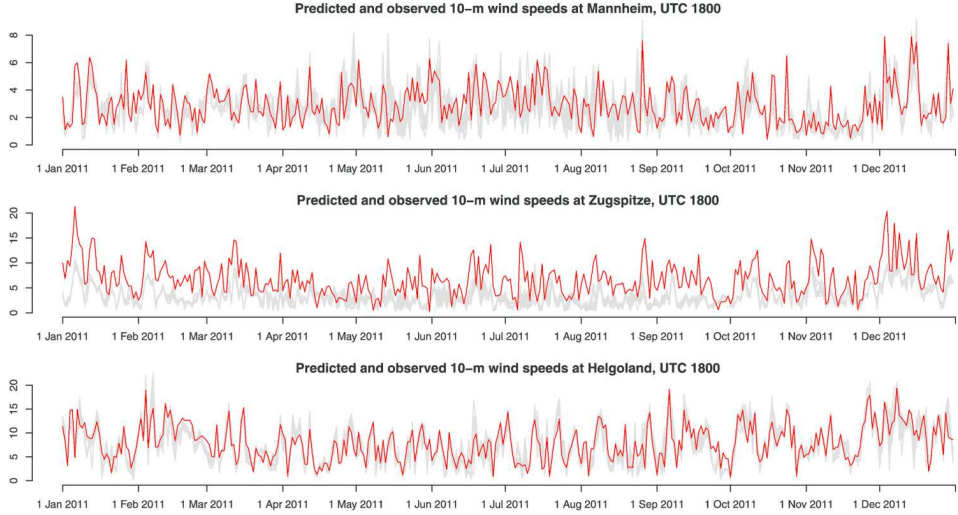}

\caption{Ensemble forecasts (light gray lines) and observations
(red lines) of wind speeds (in m/s) at 1800 UTC for all days in the
year 2011
at Mannheim, Zugspitze and Helgoland.}
\label{Fig:TimeSeries}
\end{figure}

\begin{figure}[b]

\includegraphics{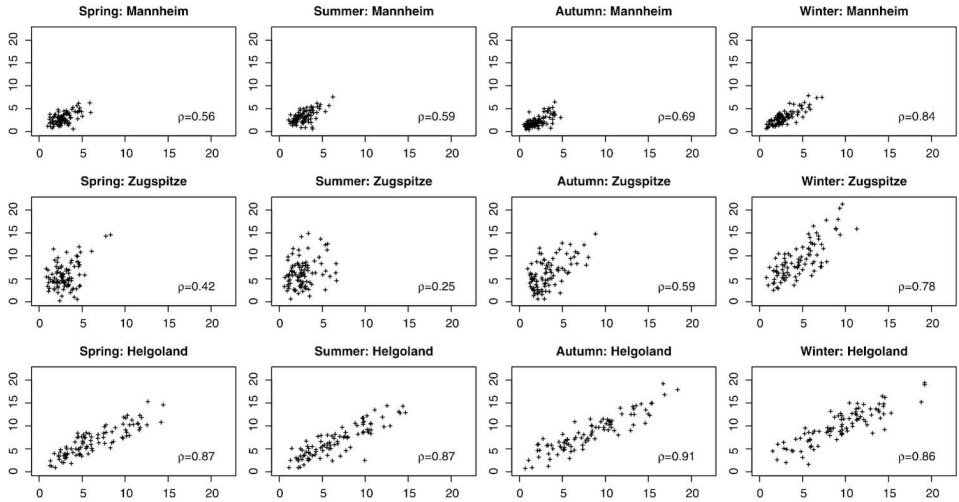}

\caption{Scatterplots of ensemble mean forecasts and
observations of wind speeds (in m/s) at 1800 UTC for all days in the
year 2011,
separately for each season.}
\label{Fig:Scatterplot}
\end{figure}

Figure~\ref{Fig:TimeSeries} shows time series of the 20 ensemble
forecasts and the corresponding observations at three different
locations in
Germany. For Mannheim, a city in southwestern Germany, and Helgoland, a small
German archipelago in the North Sea, the forecasts are generally quite accurate,
but the spread of the ensemble seems a bit low. If the ensemble
forecasts and
the observation were drawn from the same distribution, the observation
would be
contained within the ensemble range on $19/21\cdot100\%\approx90.5\%$
of all
days, which does not quite seem to be the case. The forecasts at Zugspitze,
Germany's highest mountain (located at the border to Austria), suffer
from a
systematic underforecasting bias as a result of incompletely resolved orography
by the numeric weather prediction scheme. This illustrates why a regional
post-processing approach, which assumes constant model parameters over the
entire domain of interest, is usually unable to fully remove local
biases. This
need for location-specific post-processing is further underscored by the
scatterplots in Figure~\ref{Fig:Scatterplot}, which also show that the
magnitude
of forecast error varies from one location to another. Furthermore, we note
certain differences in the predictability of wind speeds between different
seasons.\vadjust{\eject}

%

%

\section{Forecast calibration at observational sites}
\label{sec:3}

\subsection{Predictive distribution models}

For the post-processing of wind speed ensemble forecasts,
\citet{ThorarinsdottirGneiting2010} proposed an adaptation of the
nonhomogeneous Gaussian regression approach by \citet{Gneiting2005} to
nonnegative quantities, replacing normal by truncated normal predictive
distributions $\calN_0$ with a cutoff (lower bound) at zero. Specifically,
given ensemble forecasts $f_{1},\ldots,f_{m}$, they define the predictive
distribution through
%
%
\begin{equation}
\label{eq:truncnorm-model} \calN_{0}\bigl(\mu,\sigma^2\bigr),\qquad
\mbox{where } \mu=a+b_1f_1+\cdots+b_mf_m
\mbox{ and } \sigma^2=c+dS^2.
\end{equation}
Here, $S^2=\frac{1}{m}\sum_{k=1}^{m}(f_k-f_{\star})^2$ denotes the ensemble
variance, and $f_{\star}=\break \frac{1}{m}\sum_{k=1}^{m}f_k$ denotes the ensemble
mean. This type of post-processing method that fits a probability
distribution model to model output statistics (MOS) of an ensemble is also
referred to as EMOS.
Another post-processing approach based on Bayesian model averaging
[BMA, \citet{Raftery2005}] was proposed by \citet{Sloughter2010}.
In their example, gamma distributions were found to be a good model for the
conditional distribution of wind speed observations given the forecast. While
we prefer the EMOS approach over BMA due to its conceptual simplicity,
we also
study a variant of (\ref{eq:truncnorm-model}) that uses predictive gamma
distributions
%
%
\begin{equation}
\label{eq:gamma-model-mu-sigma} \calG\bigl(\mu,\sigma^2\bigr),
\qquad\mbox{where }
\mu=a+b_1f_1+\cdots+b_mf_m
\mbox{ and } \sigma^2=c+dS^2.
\end{equation}
Here, the gamma distribution is parametrized in terms of its
mean $\mu$ and variance $\sigma^2$, which relate to the shape parameter
$\alpha$ and a rate parameter $\beta$ of the standard parametrization via
$\alpha=\mu^2/\sigma^2$ and $\beta=\mu/\sigma^2$.
Yet another distribution type for wind speed observations conditional on
ensemble forecasts, the left-censored logistic distribution, has recently
been proposed by \citeauthor{Messner2014a}
(\citeyear{Messner2014a,Messner2014b}). Left-censoring a
distribution at zero entails a positive probability of observed wind
speeds being exactly zero. Truncation at zero, on the contrary, implies
that observed wind speeds can be very small, but are never exactly zero.
We favor that latter perspective and consider, as a further alternative,
an EMOS approach based on truncated logistic distributions
%
%
\begin{equation}
\label{eq:trunlogis-model} \calL_0\bigl(\mu,\sigma^2\bigr),
\qquad\mbox{where } \mu=a+b_1f_1+\cdots+b_mf_m
\mbox{ and } \sigma^2=c+dS^2.
\end{equation}
The parameters $\mu$ and $\sigma^2$ are the mean and the variance of the
logistic distribution before truncation. The common parametrization employs
a scale parameter $\varsigma$, which relates to the variance via
$\sigma^2=\frac{\pi^2}{3} \varsigma^2$. The truncated logistic
distribution resembles the truncated normal distribution but has heavier
tails (higher kurtosis). We apply all three models to wind speed forecasts
and observations directly, that is, without any prior transformation of
the data.

\subsection{Model fitting}

All of the predictive distribution models considered here depend on the
parameters $a,b_1,\ldots,b_m,c,d$ which must be estimated based on training
data. This training data usually consists of forecast-observation pairs from
the past, with the exact choice of training days depending on the weather
variable under consideration, the geographic location of the forecast domain,
etc. Temperature, for example, has a pronounced seasonal cycle, and often
also the associated forecast error statistics of the NWP model are different
in different seasons. In that case, it is therefore best to use a relatively
short rolling training period (i.e., typically around 20 to 30 days
immediately preceding the forecast day), so that the fitted parameters can
quickly adapt to seasonal changes.
The scatterplots in Figure~\ref{Fig:Scatterplot} suggest that
for 10~m wind speeds, too, the optimal model parameters may change over the
course of the year, but seasonal differences still seem to be smaller than
differences between different locations. The bias--variance trade-off that
has to be made when choosing the training sample size is therefore likely
to favor longer training periods than those typically used for temperature.
However, even if a training period of, let's say, 100 days is used for model
fitting, estimating a different set of parameters $a,b_1,\ldots,b_m,c,d$
for each location is prohibitive in the present case where we have $m=20$
nonexchangeable ensemble members and thus 23 model parameters overall. To
compromise between local adaptivity and stability of the parameter estimates,
we therefore adopt a similar approach as \citet
{ScheuererBueermann2014}, and
reparametrize our models (\ref{eq:truncnorm-model}),
(\ref{eq:gamma-model-mu-sigma}) and (\ref{eq:trunlogis-model}) in
such a
way that only three parameters are location-specific, while all remaining
parameters are assumed constant over the entire domain. Specifically, if
we denote by $\mu_s$ and $\sigma_s^2$ the mean and variance parameter of
the predictive distribution at location $s$, we let
%
%
\begin{equation}
\label{eq:re-parametrization} \mu_s = a_s + b_s(w_1f_{s1}+
\cdots+w_mf_{sm})\quad \mbox{and}\quad \sigma_s^2
= c \xi_s^2 + dS_s^2.
\end{equation}
Additive and multiplicative bias correction is controlled by the
location-specific parameters $a_s$ and $b_s$, while $w_1,\ldots,w_m$ are
nonnegative weights that are constrained to sum up to one and constant over
the entire domain. The underlying assumption is that biases vary
strongly in
space (if they are due to incompletely resolved orography, over- and
underforecasting biases may occur in close vicinity), while the relative
performance of the different ensemble members depends on the weather
situation rather than the location. In the same way, prediction uncertainty
is described by a local parameter $\xi_s^2$ that will be defined below and
two universal parameters $c$ and $d$ that control the scaling and relative
contribution of the ensemble variance $S_s^2$. Model fitting is then
performed in two steps. First, a simplified model
%
%
\begin{equation}
\label{eq:simplified-model} \mu_s = a_s + b_sf_{s\star},\qquad
\sigma_s^2 = \xi_s^2,\qquad
f_{s\star}=\frac{1}{m}\sum_{k=1}^{m}f_{sk},
\end{equation}
is fitted separately at each observation location. This model has only three
parameters $a_s, b_s, \xi_s^2$ for which reliable estimates can be obtained
even with a training data set of size 30 to 80. The estimated local
parameters are then kept fixed, data from all locations are pooled, and the
weights $w_1,\ldots,w_m$ and variances parameters $c,d$ of the
full model (\ref{eq:re-parametrization}) are estimated in a second
step. In
this step the assumption of homoscedasticity implied by
(\ref{eq:simplified-model}) is relaxed, and nonhomogeneous variances are
allowed. In contrast to generalized linear models, the variance is, however,
not related to the mean, but becomes nonhomogeneous through the use of the
additional predictor variable $S_s^2$, which provides information about the
flow-dependent forecast uncertainty.
In both model fitting steps, parameter estimation is performed as in
\citet{ThorarinsdottirGneiting2010}, that is, the model parameters are chosen
such that the corresponding predictive distributions---calculated with the
training forecasts---attain minimal continuous ranked probability score
[CRPS, \citet{Hersbach2000}] when evaluated with the training
observations. The CRPS is a proper scoring rule and can be used to rate the
sharpness and calibration of a probabilistic forecast
[\citet{GneitingRaftery2007}]. For a single predictive cumulative
distribution function $F$ and a verifying observation $y$, it is
defined as
\[
\crps(F,y) = \int_{-\infty}^\infty\bigl(F(t)-{
\mathbf1}_{[y,\infty
)}(t) \bigr)^2 \,dt .
\]
CRPS minimization is a robust alternative to maximum likelihood estimation,
which is equivalent to the minimization of the logarithmic score. For the
optimization to be computationally efficient, a closed-form expression of
the above integral must be found. For truncated normal distributions,
\citet{Gneiting2006} show that
\begin{eqnarray*}
&&\crps(F_{\calN_0(\mu,\sigma^2)},y ) \\
&&\qquad=
 \sigma\Phi\biggl(\frac{\mu}{\sigma} \biggr)^{-2} \biggl[
\frac{y-\mu}{\sigma}\Phi\biggl(\frac{\mu}{\sigma} \biggr)
\biggl\{ 2\Phi\biggl(
\frac{y-\mu}{\sigma} \biggr) + \Phi\biggl(\frac{\mu}{\sigma}
\biggr)-2 \biggr\}
\\
& & \quad\qquad{}+ 2\phi\biggl(\frac{y-\mu}{\sigma} \biggr) \Phi\biggl(\frac
{\mu}{\sigma}
\biggr)-\frac{1}{
\sqrt{\pi}}\Phi\biggl(\sqrt{2}\frac{\mu}{\sigma} \biggr)
\biggr],
\end{eqnarray*}
where $\phi$ denotes the PDF and $\Phi$ denotes the CDF of the standard
normal distribution. In Appendix~\ref{sec:A} we derive the following
expression for the CRPS of a gamma distribution with shape parameter
$\alpha$ and rate parameter $\beta$:
\begin{eqnarray*}
&&\crps(F_{\calG(\alpha,\beta)},y ) \\
&&\qquad=  y \bigl( 2 F_{\calG(\alpha,\beta)}(y)-1 \bigr) - \frac{\alpha
}{\beta} \bigl(2
F_{\calG(\alpha+1,\beta)}(y)-1 \bigr) - \frac{\alpha}{\beta\pi}
B \biggl(\alpha+
\frac{1}{2},\frac{1}{2} \biggr) ,
\end{eqnarray*}
where $B$ denotes the beta function. In Appendix~\ref{sec:B} we show
that the CRPS of a truncated logistic distribution
with location parameter $\mu$ and scale parameter $\varsigma$ is
given by
\begin{eqnarray*}
&&\crps(F_{\calL_0(\mu,\varsigma)},y ) \\
&&\qquad= (y-\mu) \biggl(\frac
{2p_y-1-p_0}{1-p_0} \biggr)
\\
& & \qquad\quad{}+ \varsigma\biggl[ \log(1-p_0) -\frac{1+2\log(1-p_y)+2p_y
\logit(p_y)}{1-p_0} -
\frac{p_0^2\log(p_0)}{(1-p_0)^2} \biggr],
\end{eqnarray*}
where $p_0=F_{\calL(\mu,\varsigma)}(0), p_y=F_{\calL(\mu,\varsigma)}(y)$,
and $\logit(p)=\log(p)-\log(1-p)$. For the minimization of the average
CRPS over all training data, we use the constrained optimization algorithm
L-BFGS-B [\citet{Byrd1995}], which allows us to enforce the constraints
$b_s,w_1,\ldots,w_m,d\geq0$ and $c>0$ for all three predictive distribution
models and the additional constraint $a_s>0$ for the gamma distribution
model.

\section{Interpolation of local predictive distributions}
\label{sec:4}

The methods described in Section~\ref{sec:3} permit location-specific
calibration of ensemble wind speed forecasts. Since both local mean and
variance parameters $\mu_s$ and $\sigma_s^2$ depend on site-specific
post-processing parameters $a_s, b_s$ and $\xi_s^2$, we face the challenge
of interpolating the local predictive distributions to nonobservational
sites such as the gridpoints of the forecast model grid. To do that,
one can
either interpolate $\mu_s$ and $\sigma_s^2$ directly, or one can interpolate
the model parameters $a_s, b_s$ and $\xi_s^2$, and use them to calculate
$\mu_s$ and $\sigma_s^2$ according to (\ref{eq:re-parametrization}).

\begin{figure}
\includegraphics{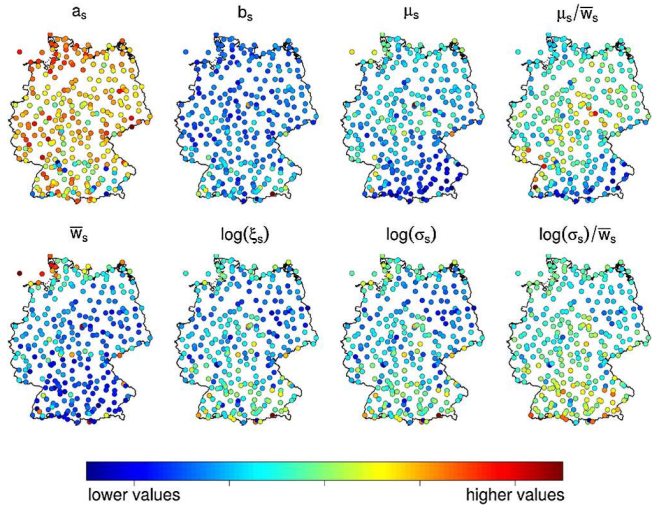}
\caption{Local post-processing parameters $a_s, b_s$ and
$\log(\xi_s)$, and parameters $\mu_s$ and $\log(\sigma_s)$ of the
predictive
distributions for wind speed at all station locations on 3 January 2012,
1800 UTC. Also shown is the average wind speed $\bar{w}_s$ at those
locations and rescaled versions of $\mu_s$ and $\log(\sigma_s)$.}
\label{Fig:ScalingIllustration}
\end{figure}

In this paper we perform spatial interpolation using a statistical
interpolation method referred to as \textit{kriging}. This technique
is based
on the assumption that the quantity to be interpolated can be
considered a
realization of a Gaussian random field (GRF), and its success depends on
whether the spatial dependence structure of this GRF is described
appropriately. Figure~\ref{Fig:ScalingIllustration} gives an idea
about the
prospective challenges with the interpolation of the different parameters
mentioned above. Although the scale and the units are different, the
plots in this figure help us identify patterns in the spatial structure of
the depicted parameters. The intercept parameter $a_s$, for example, varies
rather smoothly in Northern Germany, and spatial correlations could be
modeled well as a function of the geographical distance.
In the mountainous regions in Central and especially Southern Germany,
however, substantial small-scale variability can be observed, which makes
it rather difficult to find a reasonably simple model for spatial
dependence. Erratic small-scale departures from an otherwise smooth spatial
trend are even more pronounced with the slope parameter $b_s$.
Its interrelation with $a_s$ impedes an unambiguous
physical interpretation of these parameters. A large value of $a_s$ can be
indicative of limited forecast skill when it is accompanied by a smaller
value of $b_s$. This is the situation that we would expect for Mannheim and
Zugspitze in spring and summer (see Figure~\ref{Fig:Scatterplot}). However,
locations with similar forecast skill can still have different values
of $a_s$
if the order of magnitude of wind speeds at these locations is very different,
which is the case, for example, with Mannheim and Helgoland during the
winter season.
A large value of $b_s$ can be indicative of good forecast skill but can also
result from an underforecasting bias (e.g., at Zugspitze during winter,
see Figure~\ref{Fig:Scatterplot}).
For wind speed there is no straightforward way to reparametrize equations
(\ref{eq:truncnorm-model})--(\ref{eq:trunlogis-model}) in such a way that
bias, forecast skill and order of magnitude of the observed values can be
unambiguously attributed to different parameters. Interpolating the parameter
$\mu_s$ rather than $a_s$ and $b_s$ partly avoids this problem, but
small-scale variations are still an issue that has to be dealt with.
With $\mu_s$
having the relatively straightforward interpretation of being the expected
wind speed (up to truncation), information about local wind speed
climatologies can be used to explain regional and local differences. Such
information is available in the form of gridded annual average wind speeds
$\bar{w}_s$ over Germany in the reference period from 1981 to 2000. While
$\mu_s$ varies strongly from day to day, large values of $\mu_s$ are much
more likely to be observed at locations where wind speeds are also high on
the annual average. Indeed, the spatial patterns of $\mu_s$ and $\bar{w}_s$
in Figure~\ref{Fig:ScalingIllustration} are visually similar, and after
dividing by $\bar{w}_s$, some of the small-scale irregularities of
$\mu_s$
are strongly reduced. 
%
%
Those that are still present (or have even been
amplified by the scaling) are often observed in regions where strong
small-scale differences are present also in the annual mean, for
example, near
mountain peaks. An additional way of leveraging the information contained
in $\bar{w}_s$ is therefore to model an increase of variability between
pairs of locations not just as a function of geographic distance, but
also as a function of their difference in $\bar{w}_s$. To formalize these
ideas, we denote by $\calS$ the set of all locations within the forecast
domain and consider $\mu_s, s\in\calS$, a realization of an
intrinsic GRF
$\{Z_{\mu}(s):s\in\calS\}$ with generalized covariance function
$C_{\mu}(s,\tilde{s})$. We then study and compare several models for
spatial dependence:

\begin{longlist}[(a)]
\item[(a)] Intrinsically stationary Brownian surface plus nugget effect
\[
C_{\mu}(s,\tilde{s}) = - \theta_{\mu,1}\cdot\|s-\tilde{s}\| +
\theta_{\mu,2}\cdot\indicator{s=\tilde{s}};
\]
\item[(b)] Intrinsically stationary fractional Brownian
surface plus nugget effect
\[
C_{\mu}(s,\tilde{s}) = - \theta_{\mu,1}\cdot\|s-\tilde{s}
\|^{\theta_{\mu,3}} + \theta_{\mu,2}\cdot\indicator{s=\tilde{s}},\qquad
\theta_{\mu,3} \in(0,2);
\]
\item[(c)] Locally scaled Brownian surface plus nugget effect
\[
C_{\mu}(s,\tilde{s}) = - \theta_{\mu,1}\cdot
\bar{w}_s\cdot\bar{w}_{\tilde{s}}\cdot\|s-\tilde{s}\| +
\theta_{\mu,2}\cdot\indicator{s=\tilde{s}};
\]
\item[(d)] Locally scaled Brownian surface with an added dimension plus
nugget effect
\[
C_{\mu}(s,\tilde{s}) = - \bar{w}_s\cdot
\bar{w}_{\tilde{s}}\cdot\bigl( \theta_{\mu,1}\cdot\|s-\tilde{s}\| +
\theta_{\mu,2}\cdot|\bar{w}_s-\bar{w}_{\tilde{s}}| \bigr) +
\theta_{\mu,3}\cdot\indicator{s=\tilde{s}}.
\]
\end{longlist}
In all of these models, $\indicator{s=\tilde{s}}$ denotes the indicator
function,
$\|s-\tilde{s}\|$ is the distance between $s$ and $\tilde{s}$, and
the model
parameters $\theta_{\mu,1},\theta_{\mu,2},\theta_{\mu,3}$ are
constrained to be
nonnegative.
Model (a) is our basic model and relatively simple. It has only two parameters
which reflect the relative impact of the Brownian surface component and the
so-called \textit{nugget effect} component, which accounts for
unresolved small-scale
variability. The corresponding generalized covariance function is
\textit{conditionally
positive definite} with respect to the linear function space that
contains the
constant functions [cf. \citet{ChilesDelfiner}, or
\citet{Scheuerer2013}, for technical details on intrinsic GRFs]. It is closely
related to the exponential covariance function which has an additional
\textit{range
parameter} $r$ describing how fast correlations decay with distance. The
generalized covariance function of model (a) can be viewed as a
limiting case when $r$
tends to infinity and the variance is adjusted such that the local
characteristics
of the corresponding GRF remain unchanged. In our experiments with the
exponential
covariance, model estimates of $r$ were very large on most of the days,
and so we
decided to use the more parsimonious Brownian surface model.
The Brownian surface model (a) is a special case of the fractional Brownian
surface model (b), which contains an additional model parameter $\theta
_{\mu,3}$
that controls both fractal dimension of the realizations and growth
rate of the
variability between two locations with distance. Thus, it offers an additional
degree of flexibility, but it still assumes intrinsic stationarity of
$Z_{\mu}$,
although the above discussion of Figure~\ref{Fig:ScalingIllustration}
suggests that
this assumption is inappropriate. Our next covariance model therefore
gets back
to the idea of rendering $\mu_s$ more homogeneous by dividing it by
$\bar{w}_s$.
This kind of rescaling is equivalent to describing the spatial
correlations of
the original parameter by covariance model (c), which is again based on
the very
basic Brownian surface model but becomes nonstationary through scaling.
Note the difference in the use of covariate information compared
to \citet{ScheuererBueermann2014} who use elevation data to explain spatial
variations in temperature. In their kriging model, the covariate
information is
used to define an external drift [e.g., \citet{ChilesDelfiner},
Section~5.7.2], imposing restrictions on the kriging weights that force
them to be
consistent with the covariates. Here, the covariates are used for
rescaling the
interpolated variables, which also affects the covariance structure and accounts
for the fact that in regions where $\mu_s$ tends to be large the
magnitude of
spatial variability tends to be large as well. As a consequence of rescaling,
covariance model (c) is conditionally
positive definite with respect to the linear function space spanned by
$\bar{w}_s$ (rather than the constant functions), and this must be
taken into
account when setting up the restricted log likelihood and the kriging system
(see below).

Our second suggestion from above to leverage the information contained in
$\bar{w}_s$ in order to account for the small-scale variability of
$\mu
_s$ is
implemented
in model~(d). The two-dimensional index space $\calS$ is augmented by
a further
dimension---the average wind speed dimension---which makes it
possible to
explain large differences between locations that are geographically
close by, but
have very different wind speed climatologies. Technically, model (d)
can be thought
of as being generated by adding a separate, independent GRF (indexed
over the value
range of $\bar{w}_s$) to the 2d Brownian surface. The sum of those two
GRFs is then
rescaled with $\bar{w}_s$, and so the covariance function of model (d)
is again
conditionally positive definite with respect to the linear function
space spanned
by $\bar{w}_s$.

All of the preceding explanations concern the interpolation of the mean
parameter $\mu_s$, and we still have to specify appropriate models for
interpolating the logarithm (to ensure positivity) of the variance parameter
$\sigma_s^2$. Figure~\ref{Fig:ScalingIllustration} suggests that the
considerations
discussed above also apply to $\log(\sigma_s)$, and we therefore consider
$\log(\sigma_s), s\in\calS$, a realization of an intrinsic GRF
$\{Z_{\sigma^2}(s):s\in\calS\}$ with generalized covariance function
$C_{\sigma^2}(s,\tilde{s})$, and use the same correlation models (a)--(d) that were
discussed above for the interpolation of $\mu_s$.

%
\begin{figure}

\includegraphics{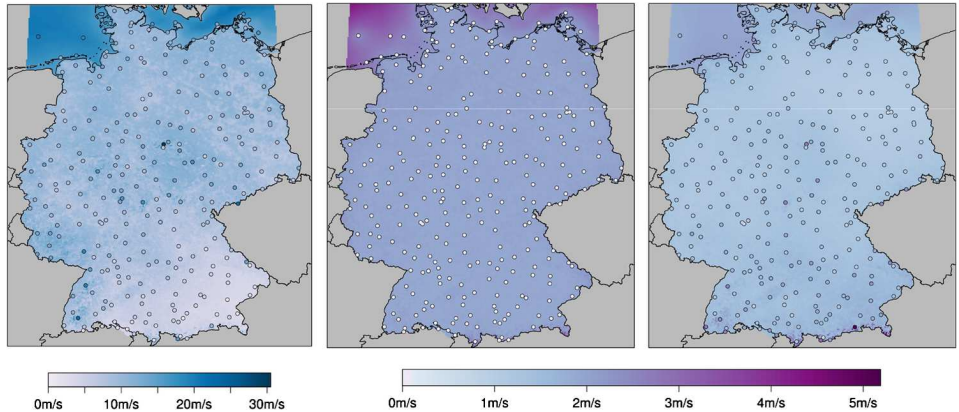}

\caption{Interpolated predictive mean $\hat{\mu}_s$ (left), corresponding
kriging standard deviation $\sigma_{\hat{\mu},s}$ (middle) and interpolated
predictive standard deviation $\hat{\sigma}_s$ (right) for wind
speeds over
Germany on January 3, 2012 at 1800 UTC.}\vspace*{6pt}
\label{Fig:InterpolationExample}
\end{figure}

Having specified the stochastic model on which we base our
interpolation scheme, we can
use standard techniques from geostatistics such as restricted maximum
likelihood (REML)
estimation and intrinsic kriging [{cf.~\citet{Scheuerer2013}, and references
therein}] to estimate the unknown model parameters and carry out the
interpolation. While
REML is based on the contestable assumption of a multivariate Gaussian
distribution of
the values of $\mu_s$ and $\log(\sigma_s)$ at the observations sites,
the discussion in
\citet{Scheuerer2013} suggests that its effectiveness does not depend
critically on this
assumption. We finally note that interpolation always involves
uncertainty, and this is
especially true in the present setting where we face a lot of
small-scale variability
that makes interpolation rather challenging. Using the interpolated
value $\hat{\mu}_s$
as the predictive mean parameter at a nonobservational location $s\in
\calS$ instead
of the unknown true value effectively increases the (interpolated) predictive
variance $\hat{\sigma}_s^2$ by the kriging variance $\sigma^2_{\hat
{\mu
},s}$. Hence, we
take the sum of those two terms as the final predictive variance
$\tilde
{\sigma}_s^2$.
The effect of the uncertainty in the interpolation of $\sigma_s^2$
itself is more
involved and would call for changing the distribution type. Within our
interpolation
scheme there is no obvious way of dealing with this appropriately, and
so we ignore this
source of uncertainty and accept its adverse effect (tails of the
predictive distribution at nonobservational sites will typically be too
light) on
forecast calibration. In Figure~\ref{Fig:InterpolationExample} we depict
the interpolated fields $\hat{\mu}_s, \sigma^2_{\hat{\mu},s}$ and
$\hat
{\sigma}_s^2$
for the 1800 UTC forecast on January 3, 2012. Those fields were
obtained with
covariance model (d) based on the parameter values of truncated logistic
distributions at the observation sites.
Owing to the covariate $\bar{w}$, the interpolation scheme can
anticipate increased
values of wind speeds and high forecast uncertainties even at locations where
the neighboring stations alone would not suggest this. Moreover, it enables
sharper transitions at the coastline than would be possible with the
basic Brownian
surface model.\vadjust{\eject}

\section{Data example}
\label{sec:5}

\subsection{Wind speed predictions at observational sites}

We first consider the situation in Section~\ref{sec:3}, where predictive
distributions are provided and evaluated at observational sites only.
The three
different distribution models are used to calibrate ensemble forecasts
of wind
speeds at 0600 UTC, 1200 UTC and 1800 UTC during the period from 1
January, 2012
to 31 December, 2012. To fit the respective model parameters, we
consider rolling
training periods of different lengths, ranging from 30 to 80 training
days. If more
than one third of the training data pairs are missing at a particular location,
no model is fitted, and the location is not considered on that
verification day.
To get a quick overview over the predictive performances of the
different methods
and different training sample sizes, we first only compare the average
CRPS over
the verification period. Recall that the CRPS is a proper scoring rule and
evaluates both calibration and sharpness of predictive distributions,
with lower
scores implying better performance.

%
\begin{table}
\tabcolsep=0pt
\caption{Average CRPS (in m/s) for the calibrated
probabilistic forecasts obtained with gamma ($\calG$), truncated normal
($\calN_0$) and truncated logistic ($\calL_0$) predictive
distribution models
and a training period of size $\mathbf{td}$. Results are given for 0600 UTC,
1200 UTC
and 1800 UTC, the corresponding CRPSs of the
raw ensemble forecasts are $0.941$, $0.960$ and $0.957$. Scores
obtained with
the simplified model (\protect\ref{eq:simplified-model}) are given in
brackets, the
best-performing approach is shown in bold}
\label{Tab:InSample}
{\fontsize{7.8}{9.8}{\selectfont
\begin{tabular*}{\textwidth}{@{\extracolsep{\fill}}lcccccc@{}}
\hline
& $\mathbf{td=30}$ & $\mathbf{td=40}$ & $\mathbf{td=50}$ & $\mathbf{td=60}$ & $\mathbf{td=70}$ & $\mathbf{td=80}$ \\
\hline
$\calG$, 0600 UTC & 0.615 (0.617) & 0.611 (0.613) & 0.609 (0.611) &
\textbf{0.608} (0.611) & 0.609 (0.611) & 0.610 (0.612) \\
$\calN_0$, 0600 UTC & 0.612 (0.614) & 0.606 (0.608) & 0.604 (0.605) &
0.603 (0.605) & \textbf{0.602} (0.604) & 0.603 (0.605) \\
$\calL_0$, 0600 UTC & 0.612 (0.614) & 0.606 (0.608) & 0.604 (0.605) &
0.603 (0.604) & \textbf{0.602} (0.604) & \textbf{0.602} (0.604) \\[3pt]
$\calG$, 1200 UTC & 0.693 (0.698) & 0.687 (0.693) & 0.684 (0.689) &
0.683 (0.689) & 0.683 (0.688) & \textbf{0.682} (0.688) \\
$\calN_0$, 1200 UTC & 0.693 (0.697) & 0.686 (0.690) & 0.681 (0.686) &
0.680 (0.685) & 0.680 (0.684) & \textbf{0.679} (0.683) \\
$\calL_0$, 1200 UTC & 0.693 (0.696) & 0.686 (0.690) & 0.681 (0.686) &
0.680 (0.684) & 0.679 (0.684) & \textbf{0.678} (0.683) \\[3pt]
$\calG$, 1800 UTC & 0.685 (0.690) & 0.680 (0.685) & 0.677 (0.682) &
\textbf{0.676} (0.681) & \textbf{0.676} (0.681) & 0.677 (0.682) \\
$\calN_0$, 1800 UTC & 0.686 (0.690) & 0.679 (0.683) & 0.675 (0.680) &
\textbf{0.674} (0.678) & \textbf{0.674} (0.678) & \textbf{0.674}
(0.679) \\
$\calL_0$, 1800 UTC & 0.685 (0.689) & 0.678 (0.683) & 0.675 (0.679) &
\textbf{0.673} (0.678) & 0.674 (0.678) & 0.674 (0.678) \\
\hline
\end{tabular*}}}
\end{table}

The results in Table~\ref{Tab:InSample} show that all post-processing
methods yield
a significant improvement in predictive performance over the raw
ensemble. The
differences between the three distribution types are very small, but
consistent over
all forecast hours and training sample sizes. They suggest that the truncated
logistic distribution yields the same or slightly better results than
the truncated
normal distribution and both are somewhat superior to predictive gamma
distributions.
In all cases, the dynamical weighting of the ensemble members and the
usage of the
ensemble variance as a predictor for the forecast uncertainty yield a slight
improvement over the simplified model (\ref{eq:simplified-model}),
which uses identical
weights for all ensemble members and assumes homoscedastic forecast errors.
The results confirm our expectations about the bias--variance
trade-off implied by the choice of the training sample size. The
optimal number of
training days is around $70$, and thus much larger than the values that
are typically
used for post-processing ensemble forecasts for temperature. Yet, it
can be observed
that initially the scores improve with increasing training sample size
due to
increasing stability of parameter estimates.
This trend is eventually reversed when further improvement in stability
becomes negligible compared to the adverse effects that come with a
reduced response
to seasonal changes. From now, we will therefore focus on the results
obtained with
a rolling training period of 70 days.

%
\begin{figure}

\includegraphics{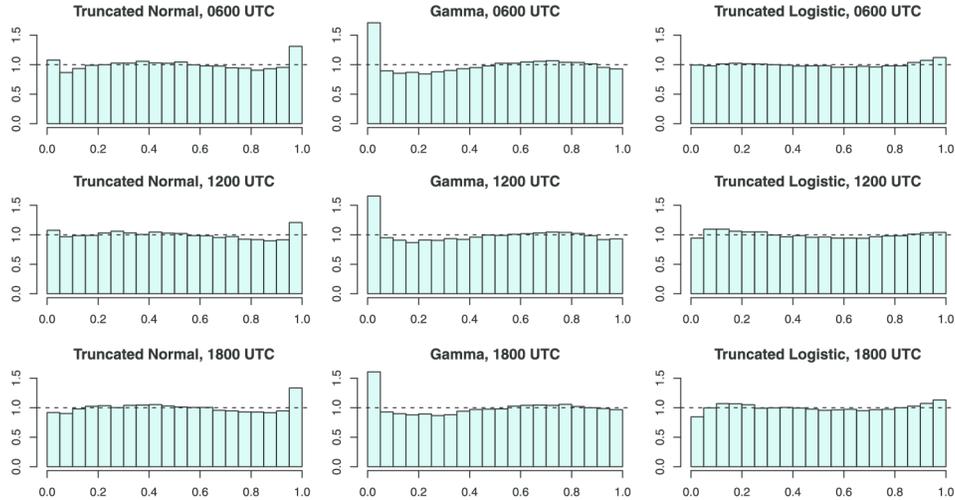}

\caption{PIT histograms for predictive gamma, truncated normal and truncated
logistic distributions at 0600 UTC, 1200 UTC and 1800 UTC (from top to bottom).}
\label{Fig:PIT-Histograms}
\end{figure}

In order to assess which distribution type yields the best calibration,
we calculate
the probability integral transforms (PITs) $\pi_i:=F_i(y_i)$ for
predictive CDFs
$F_i$ and observations $y_i$ at all locations and all verification
days. If the
forecasts are calibrated, each of those PIT values is uniformly
distributed on
$[0,1]$, and systematic departures from uniformity are indicative of a
lack of
calibration [see \citet{GneitingBalabdaouiRaftery2007} and references
therein]. Figure~\ref{Fig:PIT-Histograms} shows plots of PIT histograms
for the
three different predictive distribution models and confirms the
conclusions from Table~\ref{Tab:InSample}. All three approaches eliminate
systematic biases and give a good representation of prediction
uncertainty. However, certain differences can be observed in the tails
of those
distributions. The tails---especially the upper one---of the truncated normal
distribution model are somewhat too light. Predictive gamma
distributions, on the
contrary, give a better fit in the upper tail, but their skewness
causes the lower
tail probabilities to be too low. The truncated logistic distribution
model offers
a good compromise between the two former: it is less skewed than
the gamma distribution but has higher kurtosis than the truncated normal
distribution, thus giving an adequate fit in both lower and upper tail, and
resulting in almost perfectly flat PIT histograms.

%
\begin{figure}[b]

\includegraphics{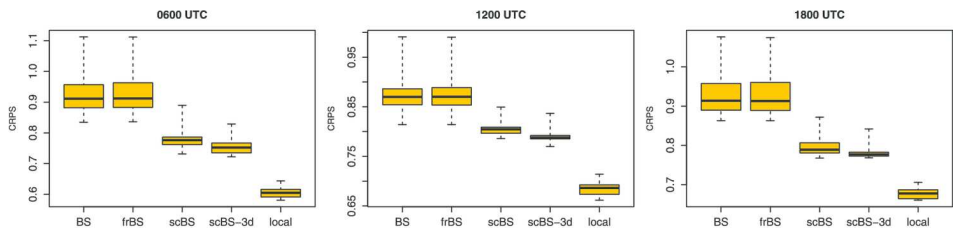}

\caption{Average CRPS values in m/s at the left-out stations obtained through
interpolation using the four different GRF models or obtained by local
calibration.}
\label{Fig:CRPS-out-of-sample}
\end{figure}

\subsection{Wind speed predictions at nonobservational sites}

We now turn to the situation where wind speed predictions are sought at
locations
where no observations are available for local calibration. In practice, those
would usually be the gridpoints of the NWP model grid. Here, in order
to be able
to measure and compare the performance of the different interpolation schemes
proposed in Section~\ref{sec:4}, we proceed as follows. From the 286 SYNOP
stations used in this study we draw 10 random samples of 50 stations
that are
left out for verification, with the sampling being done such as to
avoid clusters
of left-out stations. At the respective left-out locations, predictive
distributions (here we focus on the truncated logistic
distribution model) are obtained by interpolating the mean and variance
parameters of the retained stations, while the local observations are
used for
verification only. Again, we use the CRPS as an overall performance
measure and
compare the average CRPS values over all verification days and all left-out
stations, separately for each of the 10 different setups. In order to
see how much accuracy is lost due to the need for interpolation of mean and
variance parameters, we also give the results that are obtained when local
observations at the left-out locations are available, and $a_s, b_s$
and $\xi_s^2$
can be found as described in Section~\ref{sec:3}. The plots in
Figure~\ref{Fig:InterpolationExample} suggest that the additional
uncertainty due
to interpolation has about the same magnitude as the meteorological uncertainty
about the weather situation, which emphasizes the importance of a good
interpolation scheme. From the boxplots in Figure~\ref
{Fig:CRPS-out-of-sample} we
can see that there are substantial differences in the predictive
performance of
the probabilistic forecasts obtained with the different GRF models.
Using an
intrinsically stationary model like the Brownian surface, without
addressing the
systematic regional differences and the strong small-scale variability
of $\mu_s$
and $\sigma_s^2$, does not give an appropriate description of the
spatial dependence
structure, and entails poor interpolants of the predictive
distributions. The
fractional Brownian surface model, in spite of being more flexible, has
the same
deficiencies as the Brownian surface model and does not improve the predictions.
Using the annual mean wind speeds for locally rescaling the mean and variance
parameters, on the contrary, results in a distinctly superior
interpolation scheme,
and narrows the performance gap between the predictive distributions
obtained by
interpolation and those obtained by local calibration. The added
dimension further
improves the interpolation accuracy and yields the best predictive
performance of
all four interpolation schemes.

\subsection{Calibration of predictive means and variances}

The aim of the\break methodology proposed in Section~\ref{sec:4} is to produce
calibrated predictive distributions for wind speed at any desired location
within the forecast domain. The post-processing methods presented in
Section~\ref{sec:3} aim at adjusting predictive means and variances at
observation
locations, and Figure~\ref{Fig:PIT-Histograms} suggests that this is done
quite successfully. Do the elimination of (local) biases and the correct
representation of forecast uncertainty also carry over to locations where
no local observations are available, and predictive distributions are
obtained through interpolation? To assess this, we study again PIT values,
but we no longer pool over different locations since converse local biases
may cancel each other out. Instead, we study calibration separately for each
location and summarize the information in the PIT histograms by considering
only two statistics of the local PIT values: their mean and their mean
absolute deviation (MAD) from $0.5$. If the forecasts are calibrated,
these two quantities should be close to $0.5$ and $0.25$, respectively. A
PIT mean larger/smaller than $0.5$ is indicative for an
under-/overforecasting bias. If the forecasts are unbiased but strongly
overdispersive, the PIT values would be concentrated around $0.5$, and the
mean absolute deviations from this value would be close to zero. Conversely,
if the forecasts are strongly underdispersive, the PIT values are
concentrated near zero and one, and the MAD would be close to $0.5$. A
similar idea of reducing the information in a PIT histogram was
proposed by
\citet{KellerHense2011}, who fit a beta distribution to each histogram and
define a $\beta$-bias and a $\beta$-score which characterize the histogram
shape. The main difference is that our approach does not involve a parametric
approximation to the PIT histogram but estimates the mean and MAD of
the PIT
values directly. Apart from a reduction of information,
focusing on those two summary statistics has the advantage that meaningful
values can be calculated with relatively few PIT values and thus compensate
for the reduction of the verification sample size due to not pooling
the PIT
values over different stations.

%
\begin{figure}

\includegraphics{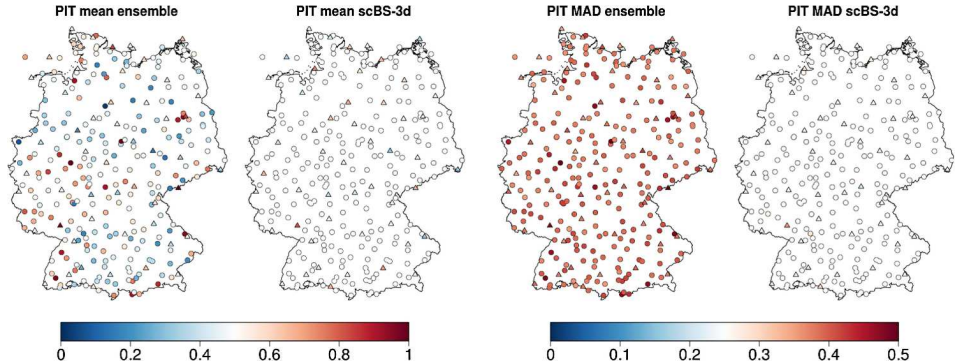}

\caption{PIT means and MADs for the raw ensemble CDFs and the
truncated logistic CDFs for wind speeds at UTC 1800, the latter being
interpolated using the scaled Brownian surface model with an added dimension.
Calibration stations are depicted as circles, left-out stations are depicted
as triangles.}
\label{Fig:PIT-maps}
\end{figure}

Figure~\ref{Fig:PIT-maps} depicts the two PIT summary statistics at
both used
and left-out stations in the first of our 10 randomly generated setups. We
compare the results for the raw ensemble and predictive truncated logistic
distributions at 1800 UTC using our best performing interpolation model (d).
The plots for the raw ensemble CDFs confirm our conclusion from the
exploratory analysis that the ensemble forecasts are strongly underdispersive,
and suffer from local biases that vary over the forecast domain. At the
observation locations, our post-processing method removes those biases
completely and yields an adequate representation of the forecast uncertainty.
At locations where predictive distributions are obtained through interpolation,
biases are mostly reduced but could not be eliminated completely, which
underscores the difficulty in calibrating forecasts in the absence of local
observations. However, our interpolation scheme is able to quantify this
interpolation uncertainty, and adding the kriging variance to the interpolated
forecast variance leads to an adequate representation of forecast uncertainty
at almost all locations of left-out stations.

\section{Discussion}
\label{sec:6}

We presented a method for post-processing ensemble forecasts of wind
speed which
can strongly improve the local calibration of raw ensemble forecasts,
even at
locations where no observations are available for calibration. Three different
types of predictive distributions---truncated normal, gamma and
truncated logistic---were studied, and were found to perform similarly in our data
example with some
advantages for the truncated logistic distribution, which turned out to
give the most adequate representation of predictive uncertainty in the tails.
In order to obtain predictive distributions at nonobservational
locations, we
used geostatistical methods to interpolate the mean and
variance parameters of the predictive distributions at surrounding observation
locations. Our results show that careful statistical modeling is
required to
formulate an adequate model for spatial dependence. In our case, the thoughtful
use of gridded data on mean annual wind speeds was a key step toward a strongly
improved interpolation scheme.

The forecasts and observations considered here were for surface wind speeds
which are relevant, for example, for severe weather warnings or airport
management. For
wind power applications, wind speeds at hub height would be more
relevant, and
our approach needs to be tested in this latter context, too. While considerably
less observations are available at hub height, the wind speed fields
are smoother
and less affected by land cover or the shape of the terrain. We expect
that our
geostatistical modeling approach could again be used successfully for generating
calibrated gridded forecasts, and we believe that the methods presented
here can
help improve, for example, the prediction of the total regional wind
energy production
based on ensemble wind speed forecasts and a few local observations.

\begin{appendix}\label{app}

\section{CRPS for gamma distributions}
\label{sec:A}

To derive a closed form expression for the CRPS of the gamma
distribution, we first
note that the CRPS can also be written as
\[
\crps(F,y) = \E_F | X-y | - \tfrac{1}{2} \E_F\bigl |
X-X^\prime\bigr|
\]
[\citet{GneitingRaftery2007}], where $X$ and $X^\prime$ are independent
random variables
with cumulative distribution function $F$ and finite first moment. For gamma
distributions $\calG(\alpha,\beta)$, the first term can be
integrated out
using the properties of their density $f_{\alpha,\beta}$, yielding
\begin{eqnarray*}
\E_{F_{\alpha,\beta}} | X-y | & = & \int_{-\infty}^y
(y-t) f_{\alpha,\beta}(t) \,dt - \int_y^{\infty} (y-t)
f_{\alpha,\beta}(t) \,dt
\\
& = & y \int_{-\infty}^y f_{\alpha,\beta}(t) \,dt -
\frac{\alpha}{\beta} \int_{-\infty}^y f_{\alpha+1,\beta}(t)
\,dt
\\
& &{} - y \int_y^{\infty} f_{\alpha,\beta}(t) \,dt +
\frac{\alpha}{\beta} \int_y^{\infty} f_{\alpha+1,\beta}(t)
\,dt
\\
& = & y \bigl( 2 F_{\alpha,\beta}(y) - 1 \bigr) - \frac{\alpha
}{\beta} \bigl( 2
F_{\alpha+1,\beta}(y) - 1 \bigr),
\end{eqnarray*}
where we have used that $\Gamma(\alpha+1)=\alpha\Gamma(\alpha)$, with
$\Gamma$
denoting the gamma function.
The second term in the above CRPS representation can be calculated by
using its
relation to the Gini concentration ratio $G$ [e.g., \citet
{McDonaldJensen1979}]:
\[
\E_{F_{\alpha,\beta}} \bigl| X-X^\prime\bigr| = \frac{2\alpha}{\beta} G =
\frac{2\alpha}{\beta} \frac{\Gamma(\alpha+{1}/{2})}{\sqrt
{\pi}
\Gamma(\alpha+1)}.
\]
Putting both terms together, replacing the fraction of gamma functions
by a beta
function and using $\Gamma(\tfrac{1}{2} )=\sqrt{\pi}$, yields
the expression
stated in Section~\ref{sec:3}.

\section{CRPS for truncated logistic distributions}
\label{sec:B}

For this calculation we take the same approach as
\citet{FriederichsThorarinsdottir2012} for generalized extreme value
distributions
and use the quantile score representation of the CRPS:
\[
\crps(F,y) = 2 \int_{0}^{F(y)} \tau
\bigl(y-F^{-1}(\tau) \bigr) \,d\tau- 2 \int_{F(y)}^1
(1-\tau) \bigl(y-F^{-1}(\tau) \bigr) \,d\tau.
\]
If we denote by $F_{\calL(\mu,\varsigma)}$ the CDF of the logistic
distribution
and let $p_0=F_{\calL(\mu,\varsigma)}(0)$, the quantile function of the
\textit{truncated} logistic distribution is given by
\[
F_{\calL_0(\mu,\varsigma)}^{-1}(\tau) = \mu+ \varsigma\logit
\bigl(p_0+\tau(1-p_0) \bigr) .
\]
After plugging this into the above quantile score representation of the CRPS
and performing integration by substitution, we obtain
\begin{eqnarray*}
&&\hspace*{-2pt}\crps(F_{\calL_0(\mu,\varsigma)},y )\\
&&\hspace*{-2pt}\qquad = (y-\mu) \biggl(\frac
{2p_y-1-p_0}{1-p_0} \biggr)
\\
& &\hspace*{-2pt}\qquad\quad{} - \frac{2\varsigma}{(1-p_0)^2} \int_{p_0}^{p_y} (
\tau-p_0) \logit(\tau) \,d\tau+ \frac{2\varsigma}{(1-p_0)^2} \int
_{p_y}^1 (1-\tau) \logit(\tau) \,d\tau,
\end{eqnarray*}
where $p_y=F_{\calL(\mu,\varsigma)}(y)$. The two integrals can be calculated
using
\begin{eqnarray*}
2 \int(\tau-p_0) \logit(\tau) \,d\tau& = & \bigl(
\tau^2-2p_0\tau\bigr) \logit(\tau) +
(1-2p_0) \log(1-\tau) + \tau,
\\
2 \int(1-\tau) \logit(\tau) \,d\tau& = & - (1-\tau)^2 \logit
(\tau) +
\log(\tau) - \tau,
\end{eqnarray*}
and after some rearrangement and simplification we finally obtain the
expression stated in Section~\ref{sec:3}.
\end{appendix}
\section*{Acknowledgments}
The authors are grateful to Tilmann Gneiting for helpful comments. They thank
Sabrina Bentzien and all members of the COSMO-DE-EPS project at
Deutscher Wetterdienst
for their support with the acquisition of the ensemble forecast data.

%





\printaddresses
\end{document}